# Polariton Enhanced IR Reflection Spectra of Epitaxial Graphene on SiC

B.K. Daas[*], K. M. Daniels, T.S. Sudarshan, M.V.S. Chandrashekhar

Department of Electrical and Computer Engineering, University of South Carolina
301 S. Main St, Columbia, SC 29208, USA
*Daas@email.sc.edu

***Abstract:***

We show ~10x polariton-enhanced infrared reflectivity of epitaxial graphene on 4H-SiC, in SiC's restrahlen band (8-10µm). By fitting measurements to theory, we extract the thickness, N, in monolayers (ML), momentum scattering time ($\tau$), Fermi level position ($E_F$) of graphene and estimate carrier mobility. By showing that $\tau \propto 1/\sqrt{n_s}$, the carrier concentration/ML, we argue that scattering is dominated by short-range interactions at the SiC/graphene interface. Polariton formation finds application in near-field optical devices such as superlenses.

Graphene, a two-dimensional (2D) form of carbon in a honeycomb crystal structure, is the basic building block of other $sp^2$ carbon nanomaterials, such as carbon nanotubes. It exhibits unusual electronic and optical properties [1-6]. Graphene has a dispersionless linear electronic band structure as opposed to the quadratic form observed for most semiconductors. This leads to "massless" Dirac-fermion behavior, and consequently, high electron mobility, as opposed to the usual Schrodinger behavior exhibited by most semiconductors [6-7]. Furthermore, the recent development of epitaxial graphene (EG) formed by the solid-state decomposition of a SiC surface has enabled the systematic production of large area graphene films on a commercial substrate platform. This has prompted the investigation of many high performance electronic devices, such as field effect transistors and p-n junction diodes, photonic devices such as terahertz oscillators, as well as low noise sensors [8-11]. For all of these applications, knowledge of the optical properties of graphene is important, as it gives insight into the interaction of graphene with external electromagnetic fields.

Theoretical calculations of the optical conductivity of graphene, using the Dirac Hamiltonian were performed by Gusynin and Sharapov [12] and others [13]. A more detailed study of optical conductivity by Stauber et al. [14] analyzes the reflectance and transmittance of a graphene plane located at the interface of two different dielectrics, a practical measurement geometry. These calculations were confirmed by measurements of IR-reflectivity and transmission of exfoliated graphene on glass substrates. [15].

Within this theoretical framework, the IR transmission spectra of EG on semi-insulating SiC substrates were studied by Dawlaty et al [16], where thickness, N, in monolayers (ML) ,$\tau$ the momentum scattering time, and $|E_F|$, the Fermi level position in EG were extracted by fitting measurement with theory. However, the substrate dielectric function used was a constant value (~6.5), independent of wavelength. This is inaccurate for SiC, particularly in the restrahlen band (8-10µm range), where there is a singularity in the dielectric function, leading to negative dielectric function in this range. Such anomalous dielectric behavior in SiC has been well-documented and predicts the formation of a polariton at the SiC/EG interface [17]. A polariton is a bosonic quasi-particle comprising of a photon coupled with a transverse electromagnetic field [18]. In SiC, the polariton originates from the strong coupling of the transverse electromagnetic field in the incident light with the optical phonons in the substrate (~800-1000cm$^{-1}$), damping the electromagnetic wave in the substrate. This is the phenomenon responsible for the singularity in the dielectric function. The formation of the polariton in the restrahlen band results in decreased transmittance and increased reflectance compared to outside the restrahlen band, potentially allowing more precise determination of N.

Determination of N for EG is challenging as 1ML is only 0.33nm thick. Atomic force microscopy (AFM) in contact mode is the most direct way to obtain N, but it is slow and may damage the crystalline lattice during measurement. Furthermore AFM instrumental offset can be as high as ~0.5nm, thicker than a single graphene layer, resulting in potentially ambiguous measurements [19]. Raman spectroscopy is a convenient nondestructive method for quick inspection of graphene thickness [19-21]. However, Raman does not give unambiguous measurements of N with ML precision. Reflection and contrast spectroscopy have been used to precisely determine graphene thickness on SiO$_2$ substrates [22]. It has been shown that changes in optical response in reflection, induced by the presence of graphene are larger than for transmission [15]. When graphene is of thickness d≪ λ, the wavelength of light, the solution to Maxwell's equation indicates fractional change is reflectance $\propto \frac{4}{n_2^2-1}$ compared to transmission change $\propto \frac{2}{n_2+1}$, where $n_2$ is the refractive index of SiC. This gives ~20% greater response in differential reflection than transmission for graphene on SiC substrates.

While IR transmission measurements have been performed on EG [16], IR reflection measurements have not, despite the promise of greater accuracy. Although Dawlaty et. al [16] were able to determine N for EG on SiC using IR-transmission, there was still uncertainty in the thickness thus obtained, as there was discrepancy between IR-thickness values and those determined by other techniques such as Raman spectroscopy and x-ray photoelectron spectroscopy (XPS), along with potential issues with removal of unintentionally grown graphene on the backside. Therefore, in this paper, we present the reflection spectra of EG on N+ 4H-SiC substrate from mid IR (~12 THz) to IR (~120 THz) frequency regime. A detailed mathematical model taking into account the full SiC dielectric function was fit with the experimental results to extract N, τ and IE$_F$I while clearly demonstrating the formation of the polariton at the SiC/EG interface.

The optical conductivity of graphene both in interband (σ$_{inter}$) and intraband (σ$_{intra}$) is given by [16]

$$\sigma_{inter}(\omega) = i\frac{e^2\omega}{\pi}\int_\Delta^\infty d\mathcal{E}\frac{(1+\frac{\Delta^2}{\mathcal{E}^2})}{(2\mathcal{E})^2-(\hbar\omega+i\Gamma)^2}[f(\mathcal{E}-E_F)-f(-\mathcal{E}-E_F)] \quad (1)$$

$$\sigma_{intra}(\omega) = i\frac{\frac{e^2}{\pi\hbar^2}}{\omega+\frac{i}{\tau}}\int_\Delta^\infty d\mathcal{E}(1+\frac{\Delta^2}{\mathcal{E}^2})[f(\mathcal{E}-E_F)+f(\mathcal{E}+E_F)] \quad (2)$$

Here $f(\mathcal{E}-E_F)$ is the Fermi distribution function with the Fermi energy $E_F$, Γ describes the broadening of the interband transition [16] and 2Δ accounts for a potential band gap in graphene. We used Δ= 0 and Γ=10meV, as has been used by other researchers [16]. At high frequencies, the real part of σ$_{inter}$(ω) attains a constant value. At low frequencies, it approaches zero, as interband optical transition are blocked due to the presence of electrons and holes near the band edges [16]. The plasmon dispersion and free-carrier absorption are described by σ<sub>intra</sub> (ω), similar to the Drude model, with band occupancy accounted for. The total conductivity σ(ω), is the sum of σ<sub>inter</sub>(ω) and σ<sub>intra</sub>(ω).

Considering graphene at the interface between two dielectrics with dielectric functions $\varepsilon 1$ and $\varepsilon 2$, the total reflectivity is given by[14]

$$R = = \frac{|(\sqrt{(\varepsilon 1 \varepsilon 2 \varepsilon 0)}/\alpha) + \frac{\sqrt{\varepsilon 1} N\sigma(\omega) \times \cos(\Theta_I)}{c} - \varepsilon 1 \varepsilon 0|^2}{|(\sqrt{(\varepsilon 1 \varepsilon 2 \varepsilon 0)}/\alpha) + \frac{\sqrt{\varepsilon 1} N\sigma(\omega) \times \cos(\Theta_I)}{c} + \varepsilon 1 \varepsilon 0|^2} \quad (3)$$

Where $\alpha$ is given as $\alpha = \frac{\sqrt{1-[(\frac{n1}{n2})\sin\Theta_I]^2}}{cos\Theta_I}$ n1 and n2 are the refractive index of air and SiC respectively, σ(ω) is the total conductivity, $\varepsilon 0$ is the free space permittivity (~8.854×10<sup>-12</sup> F/m). For EG on SiC substrates, $\varepsilon 1$ is the permittivity of air (~1) and $\varepsilon 2$ is the permittivity of SiC, which is a function of wavelength, given by [17]

$$\varepsilon 2 = \varepsilon 2(\omega) = \varepsilon_\infty \frac{\omega^2 - \omega_{LO}^2 + i\Gamma_1 \omega}{\omega^2 + \omega_{TO}^2 + i\Gamma_2 \omega} \quad (4)$$

Here $\varepsilon_\infty$ =6.5 is the positive ion core background dielectric constant, $\omega_{LO}$ is the longitudinal optical phonon frequency ($\omega_{LO}$=972cm<sup>-1</sup>) , $\omega_{TO}$ is the transverse optical phonon frequency ($\omega_{LO}$=796 cm<sup>-1</sup>). Γ<sub>1,2</sub> describes the broadening of the phonon resonances, typically 5-60 cm<sup>-1</sup>, where the higher values are due to free-carrier absorption . By measuring the IR reflectivity of N+ 4H-SiC substrates, we found Γ<sub>1</sub>=60cm<sup>-1</sup> while Γ<sub>2</sub>=10cm<sup>-1</sup>. Γ<sub>1,2</sub> were used as free fitting parameters in the measurement of the EG/SiC interface, with these nominal values for Γ<sub>1,2</sub> as the starting point.

Figure 1 shows the theoretical prediction of the variation of the differential reflectance spectra of EG, with respect to a SiC substrate, as a function of N, $E_F$ and τ respectively. Figure 1(a) shows that reflectivity increases as N increases. Increasing N gives higher conductivity (equation 3), leading to enhanced interaction with the incoming light. This leads to increased reflectivity similar to the Drude model [23]. The formation of the polariton can clearly be observed as a reflectivity hump in the frequency region between the pole at ω=$\omega_{TO}$ and the zero at ω=$\omega_{LO}$ (equation.4), the width of the restrahlen band for which $\varepsilon 2(\omega)$ is negative. This polariton manifests as the enhanced reflectivity in this regime.

Figs 1(b) and 1(c) illustrate the variation of reflectivity with |$E_F$| and τ. These variations change the hump shape and location in the 5-7um region leading to the restrahlen band. This is due to the transition from intraband to interband optical conductivity as wavelength changes. The change of |E<sub>F</sub>| and τ changes the position of this transition, as has been observed before [16]. Increasing |E<sub>F</sub>| and decreasing τ also decrease the peak reflection spectra in the restrahlen band, due to greater free carrier absorption. We are unable to find the exact sign of |E<sub>F</sub>| because of electron-hole symmetry of the EG band structure near the band edge near the K point in the Brillouin zone of graphene [5]. Thus, this model is equivalent for p and n-type EG.

Epitaxial growth of large-area graphene by thermal decomposition of commercial <0001> 4H and 6H SiC substrates at high temperature and vacuum has been demonstrated [6]. This produces EG a few ML to >50 ML thick, depending on growth conditions. In our experiments, EG was grown on commercial N+ 4H-SiC substrates, nitrogen doped ~$10^{19}$/cm$^3$. 1cmx1cm samples were degreased using Trichloroethylene (TCE), acetone and methanol respectively. They were then rinsed in DI water for three minutes. The samples were finally dipped in HF for two minutes to remove native oxide and rinsed with DI water before being blown dry. They were then set in the crucible in an inductively heated furnace where high vacuum was maintained (<$10^{-6}$ Torr) and baked out at 1000$^0$C for 13 to 15 hours. The temperature was slowly raised to the growth temperature (1250-1400°C). All growths were performed for 60 minutes before cooling to 1000$^0$C at a ramp rate of 7~8$^0$C/min and eventually to room temperature. Slow temperature ramps were utilized to minimize thermal stress on the samples.

After growth AFM (atomic force microscopy) and Raman measurements were carried out on EG on both carbon (C) and silicon (Si) faces. Micro-raman spectroscopy using a 632nm laser shows the G peak (~1590cm$^{-1}$), D peak (~1350cm$^{-1}$) and 2D peak (~2700cm$^{-1}$) characteristic of EG [24]. The ratio of intensities of the D-peak to G-peak, $I_D/I_G$ **≤0.2** demonstrates the quality of our graphene [24]. All Raman spectra in this paper have the SiC substrate signal subtracted i.e. only difference Raman spectra of the EG are presented[24]. Fourier transform infrared (FTIR) reflectivity measurements were performed using a Galaxy Series FTIR-5000 spectrometer in an incidence angle of 40$^0$ over the wavelength 2.5μm to 25μm using a blank N+ SiC substrate as the reference. Measured FTIR data was fit with the mathematical model to extract N, τ and $E_f$.

X-ray photoelectron spectroscopy (XPS) measurements were conducted using a Kratos AXIS Ultra DLD XPS system equipped with a monochromatic Al Kα source. The energy scale of the system was calibrated using a Au foil with Au4f scanned for the Al radiation and a Cu foil with Cu2p scanned for Mg radiation resulting in a difference of 1081.70 ± 0.025eV between these two peaks. The binding energy is calibrated using a Ag foil with Ag3d$_{5/2}$ set at 368.21 ± 0.025eV for the monochromatic Al X-ray source. The monochromatic Al Kα source was operated at 15 keV and 150 W. The pass energy was fixed at 40 eV for the detailed scans. A charge neutralizer (CN) was used to compensate for the surface charge. Using XPS we obtained C 1s peak and Si2p peak both on the EG samples and on the substrate, in normal and 70$^0$ beam incidence angles to overcome any instrumental error. Graphene C1s peak was normalized to the SiC substrate C1s peak, from which the thickness was determined as described elsewhere [24-25]. The thickness N obtained using FTIR and XPS was consistent to within 2ML.

Figure 2 illustrates two samples with reflectance spectra along with the corresponding AFM images. Fig 2(a) shows an EG layer on n+ Si-face SiC with N=2, while Fig 2(b) shows a bare SiC substrate with no EG. Difference Raman spectra of the of G peak, D peak and 2D peak are found at 1596cm$^{-1}$, 1354cm$^{-1}$ and 2694cm$^{-1}$ respectively, indicating the presence of graphene[24].The IR reflectivity spectra are normalized to the bare SiC substrate shown in Figure 2b).

Comparing the IR-reflectivity of EG with SiC, we clearly observe the hump corresponding to the polariton at the EG/SiC interface, as discussed earlier. For even a very thin layer e.g. N=2, (Fig 2a(I)), the reflectivity is ~20% higher than bare SiC in the restrahlen band, which is remarkable compared to the increased reflectivity of (2% to 3% greater than bare SiC) graphene outside the restrahlen band. This strong effect is unique to the SiC/graphene system, and may have important applications in EG/SiC based plasmonics, and will be discussed at the end of the paper.

Fig 3 shows normalized reflection spectra matched with the theoretical model for N=2 (Si-face), 9 (C-face) and 17 (C-face) ML's for three different samples respectively, using equation (1) to (4) to extract N, τ and $|E_F|$. We observe good correlation between experiment and theory. No significant differences were observed between the Si-face and C-face in terms of IR-reflectivity. As expected, the total reflectivity increases with N.

Fig 4(a) shows the variation of $|E_F|$ with N, where $|E_F|$ was found to decrease with increasing N. This is in agreement with electrostatic screening length of ~1ML [29], which means that only the first monolayer near the substrate is doped, while the remaining are intrinsic. We note that we use a single $|E_F|$ to describe all the layers, representing an average over all the layers, leading to total conductivity (Nσ(ω)) of the EG layer. The averaging process gives an effective screening length somewhat greater than 1 ML, as it quickly reduces the effective $|E_F|$ for thin layers N<3.

From the obtained $|E_F|$, the graphene carrier density per ML, $n_s$, can be calculated using [26]

$$n_s = \int_0^\infty D(E) f(E - E_F) \, dE \qquad (5)$$

where the density of states for graphene, $D(E) = 2E/\pi(\hbar v_F)^2 \approx 1.46 \times 10^{14}$ E cm$^{-2}$ and $f(E)$ is the Fermi distribution function dependent on $|E_F|$. As $|E_F|$ extracted from our experiment represents an average over all the layers, $n_s$ represents the average carrier density per ML. Fig 4(b) shows the variation of τ with $n_s$ where the thickness corresponding to each data point is indicated. The distance between two electrons is $\sqrt{\frac{1}{n_s}}$ while, $v_F$ is the carrier fermi velocity. Thus, the time, τ, between scattering events can be obtained,

$$\tau = k_1 \left(\sqrt{\frac{1}{n_s}}\right) / v_F \qquad (6)$$

where $k_1$ is a "fitting factor" constant of order unity, which accounts for other short-range scattering mechanisms [27-28], such as from lattice defects, as well as geometrical effects. Our experimental values were fit very well with equation (6) using $k_1$=0.6 (shown in fig.4(b)), which is not significantly different from unity. Our measured τ values are similar to those extracted in [16]. Discrepancies might be attributed to the use of N+ 4H-SiC substrates (our work) versus the use of insulating SiC substrates [16], along with uncertainties in backside graphene removal in transmission [16].

Electronic transport in graphene is dominated primarily by two scattering mechanisms: i) short-range scattering and ii) long-range coulomb scattering [28]. Short range scattering originates from short-range factors such as lattice defects and electron-electron interactions. It is temperature independent and leads to $\tau \propto \sqrt{\frac{1}{n_s}}$, as discussed above. Long range scattering, on the other hand, originates from the screening of charged impurities on the surface of graphene, and gives $\tau \propto \sqrt{n_s}$. Clearly, these two dramatically different mechanisms can be distinguished from Figure 4(b).

The fact that we can describe the optical conductivity of our samples so accurately with this $\sqrt{\frac{1}{n_s}}$ dependence (equation (6)) strongly indicates that we are dominated by short-range scattering, rather than by long-range coulomb scattering. This conclusion also suggests that our samples have little surface contamination. Short-range scattering is temperature independent. Therefore, the temperature dependence of τ must be investigated to conclusively determine the nature of the scattering in these measurements. Nevertheless, these measurements demonstrate that EG might intrinsically approach the behavior of ideal graphene, as opposed to exfoliated graphene, whose behavior is dominated by long-range impurity scattering [27-28]. Furthermore, our measurements are performed at optical frequencies, where the conductivity behavior may quantitatively be very different from at low frequencies (DC) [16]. The reconciliation of the physics in the two frequency regimes bears further investigation.

Using these parameters, the electron mobility can be estimated using $e\tau v_f^2/E_f$ [16]. This gives ~1000cm$^2$/Vs for Si-face thin layers, and ~10000cm$^2$/Vs for thicker layers on the C-face, in agreement with DC transport measurements in the literature[29]. In other words, as $n_s$/ML decreases, mobility increases. This technique may enable non-destructive characterization of the electronic properties of EG, although further work on correlating DC transport measurements with FTIR is required.

The polariton enhanced reflectivity of EG can be used in many nanophotonic device and sensor applications. The polariton at SiC/EG interface can be used to overcome the diffraction limit of light, leading to on-chip nanophotonic devices [30]. In these applications, light is converted to a polariton at SiC/EG interface. This polariton exhibits much shorter wavelength than a traveling electromagnetic wave, giving a much smaller footprint on-chip. This optical signal can then be processed as a polariton according to the application requirements, at the end of which the polariton is converted back to light for retransmission. In other words, the EG may serve as an efficient plasmonic waveguide. The overcoming of the diffraction limit also enables the realization of near-field superlenses [17]. This characteristic may also be used as bio molecular or gas sensor where adsorption doping occurs [31], changing |$E_F$|, resulting in a change of reflectivity. Other novel nanophotonic devices such as cloaks [32] and Bragg gratings [30] may be enabled by this phenomenon.

In summary, we have demonstrated that by using IR-reflection spectra, we can conveniently determine N, τ and |$E_F$| for EG grown on SiC substrates, with N-measurements correlated with XPS to within 2ML. These characteristics were obtained by fitting to a simple optical conductivity model based on the Dirac Hamiltonian. Excellent correlation was observed between theory and experiment. The model worked equally well for C and Si-face-grown EG. An analysis of these parameters indicated that short-range scattering was dominant in our layers, showing that these layers behave like intrinsic graphene (no coulomb-scattering), at least in the IR-regime. The observation of significantly enhanced reflectivity in SiC's restrahlen band demonstrates the formation of a polariton at the SiC/EG interface, as predicted by theory from our model. This effect enables sub-wavelength near field optical applications using EG/SiC.

The authors would like to acknowledge the Southeastern Center for Electrical Engineering Education, and the National Science Foundation for funding this work. They would also like to thank "Arva Hindu Foundation" along with Professor Yogendra P. Kakad at UNCC for partial financial support of this work. They also gratefully acknowledge Dr. Chris Williams in Chemical Engineering at University of South Carolina for the use of his Raman spectrometer.

**References:**

[1] R.Saito, G. Dresselhaus, and M.S. dresselhaus; Phy.Rev.B 61, 2981 (2000)

[2] A.H. Castro Neto, F Guinea, N.M.R Peres, K.S. Novoselov and A.K. Geim; Rev. Mod. Phys. 81, 109 (2009)

[3] K. S. Novoselov, A. K. Geim, S. V. Morozov, D. Jiang, M. I. Katsnelson, I. V. Grigorieva, S. V. Dubonos, and A. A. Firsov, Nature-London, 438, 197-201 ,2005.

[4] Wallace, P.R. Phys. Rev. 71, 9, 622 (1947)
[5] Y. Zhang, Y. Tan, H. L. Stormer, and P. Kim; Nature Lett, 438, 201 (2005).
[6] C. Berger, Z. Song, X. Li, X. Wu, N. Brown, C. Naud, D. Mayou, T. Li, J. Hass, A. N. Marchenkov, E. H. Conrad, P. N. First, and W. A. de Heer; J. Phys. Chem. B 108, 19912 (2004)
[7] Haldane, F.M.D; Phys. Rev. Lett. 61,18, 2015 (1988)
[8] G. Liang, N. Neophytou, D. E. Nikonov, and M. S. Lundstrom, IEEE Trans. Electron Devices 54, 677 (2007)
[9] J. R. Williams, L. DiCarlo, and C. M. Marcus; Science, 317, 638-641 (2007)
[10] G. Gu, S. Nie, R. M. Feenstra, R. P. Devaty, W. J. Choyke, W. K. Chan, and M. G. Kane, Appl. Phys. Lett. 90, 253507 (200)
[11] F. Rana, IEEE Trans. Nanotechnol. **7**, 91 (2008)
[12] V.P. Gusynin, S.G. Sharapov; Phys. Rev. B 73, 245411 (2006)
[13] V. P. Gusynin, S. G. Sharapov, and J. P. Carbotte; Phys. Rev. B 75, 165407 (2007).
[14] T. Stauber, N.M.R Peres, A.K. Geim; Phy.Rev. B 78 085432 (2008)
[15] Kin Fai Mak, Matthew Y. Sfeir, Yang Wu, Chun Hung Lui, James A. Misewich and Tony F. Heinz, Phy. Rev. Lett. 101, 196405 (2008)
[16] Jahan M. Dawlaty, Shiriram Shivaraman, Jared Strait, Paul George, Mvs Chandrashekhar, Farhan Rana, Michael G. Spencer, Dmity veksler and Yunqing Chen, Appl. Phys. Lett. 93,131905 (2008)
[17] Dmitriy Korobkin, Yaroslav Urzhumov, and Gennady Shvets; J. Opt. Soc. Am. B, 23,3,468 (2006)
[18] Charles Kittel "Introduction to Solid State Physics" 8$^{th}$ edition, John Wiley & sons,Inc, ISBN 0-471-41526-X, (2005)
[19] Gupta, A; Chen,G.;Joshi,P.;Tadigadpa, S.; Eklund, P.C.; Nano Lett. 6,12, 2667 (2006)
[20] Ferrari, A.C.; Meyer, J.C; Scardaci, V.; Casiraghi, C.; Lazzeri, M.; Mauri, F.;Piscanec, S.; jiang, D.; Novoselov, K.S.; Roth, S.; Geim, A.K. Phys. Rev. Lett. 7, 238 (2006)
[21] Graf, D.; Molitor, F.; Ensslin, K.; Stampfer, C.; Jungen, A.; Hierold, C.; Wirtz, L.; Nano Lett.,7,2, 238 (2007)
[22] Z.H.Ni, H.M. Wang, J. Kasim, H.M. fan, T. Yu, Y.H. Wu, Y.P. Feng and Z.X. Shen; Nano Lett..I.7,9 2758 (2007)
[23] T.-I. Jeon and D. Grischkowsky, Appl. Phys. Lett. 72,23, 3032 (1998)
[24] Shriram Shivaraman, M.V.S. Chandrashekhar, Jhon J. Boeckl and Michael Spencer; J. of Elec. Matt., 38,6 (2009)
[25] P.J.Cumpson; Surf.Interface.Anal,29,403 (2000)
[26] Low Bias Transport in Graphene: An Introduction, *2009* NCN@Purdue Summer School: Electronics from the Bottom Up.
[27] E.H. Hwang, S. Adam and S. Das Sarma; Phy.. Rev. Lett. 98 106806 (2007)
[28] S. Adam, E.H. Hwang, S. Das Sarma; Physica E 40 1022 (2008)
[29] Walt A de Heer, Claire Berger, Xiaosong Wu, Mike Sprinkle, Yike Hu, Ming Ruan, Joseph A Stroscio, Phillip N First, Robert Haddon, Benjamin Piot, Clément Faugeras, Marek Potemski and Jeong-Sun Moon;J.Phys.D:Appl.Phys. 43 374007 (2010)
[30] M. Dragoman, D. Dragoman; Progress in Quantum Electronics 32, 1-41 (2008)
[31] Muhammad Qazi, Mohammad W.K. Nomani, M.V.S. Chandrashekhar, Viragil B. Shields, Michael G. Spencer and Goutam Koley; Appl. Phys. Express 3, 075101 (2010)
[32] Wenshan Cai, Uday K. Chettiar, Alexander V. Kildishev, and Vladimir M. Shalaev; Optics Express,16,8,5444-5452 (2008)

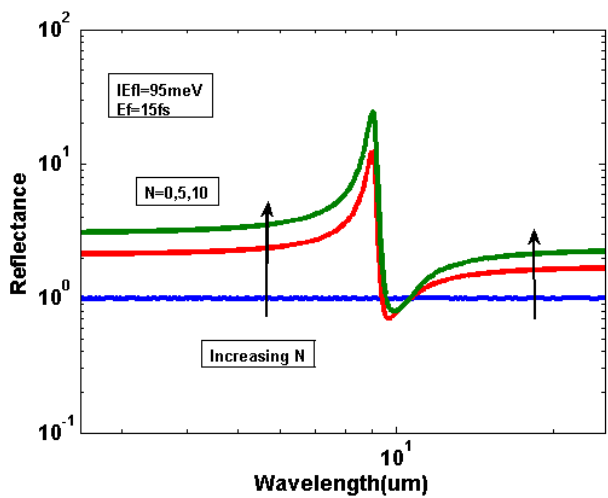

Fig1(a): Variation of Reflection of graphene for various N where |E$_F$|=95meV and τ=15fs.

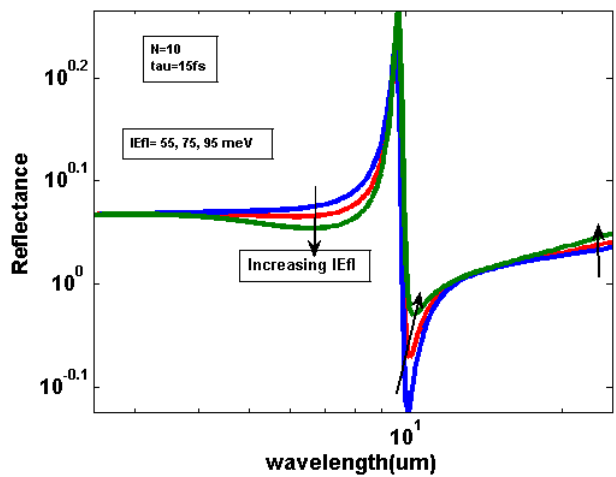

Fig1(b): Variation of Reflection of graphene for various |E$_F$| where N=10 and τ=15fs.

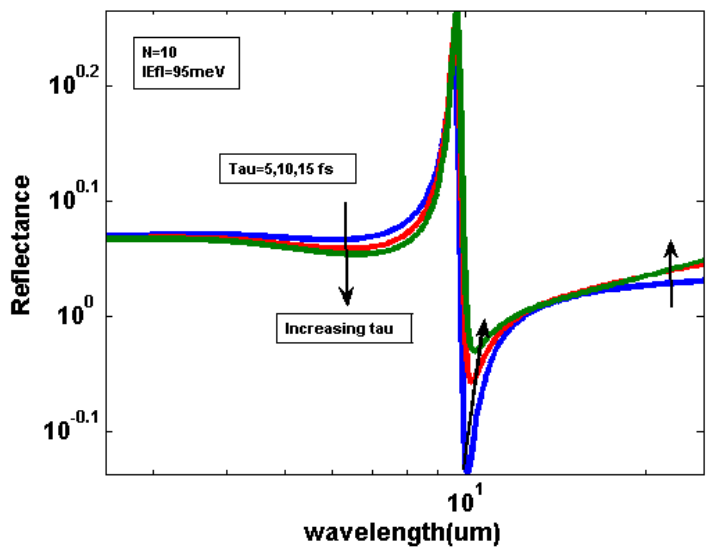

Fig1(c): Variation of Reflection of graphene for various τ where N= 10 and |E$_F$|=95meV.

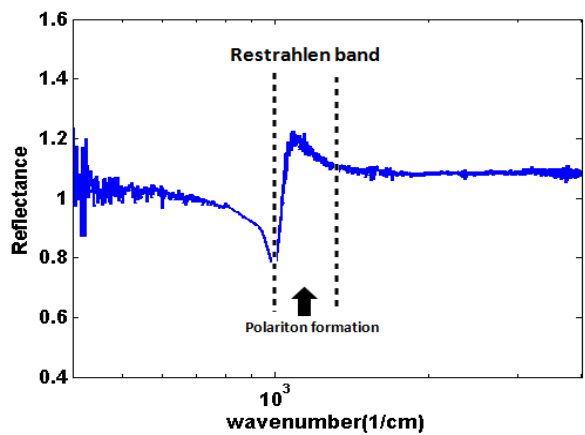

Fig:2a(I): Reflection spectrum of EG (2 ML)

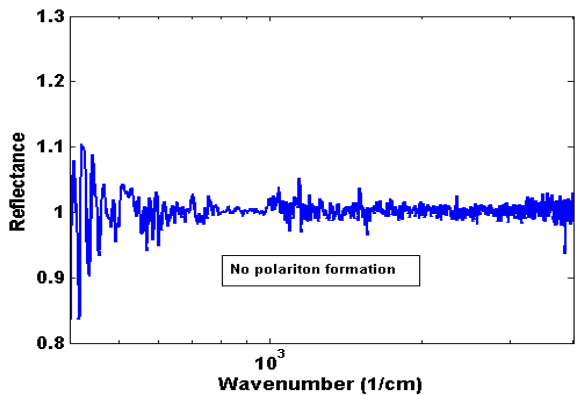

Fig2b(I): Reflection spectrum of SiC substrate

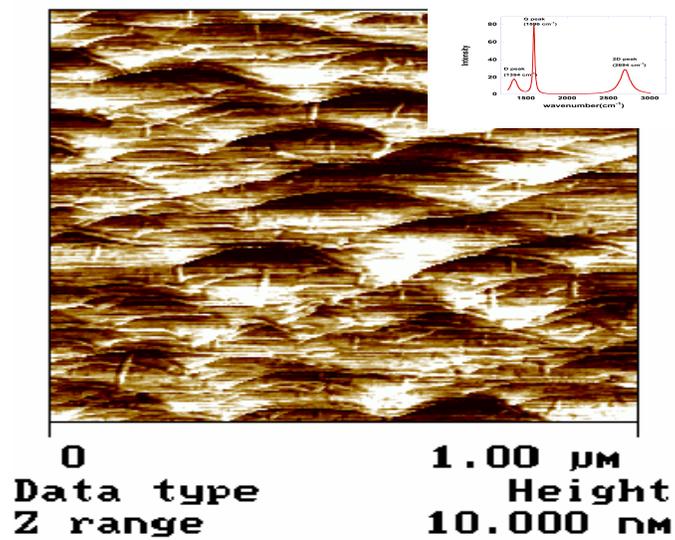

Fig2a(II): AFM image of EG (2ML)

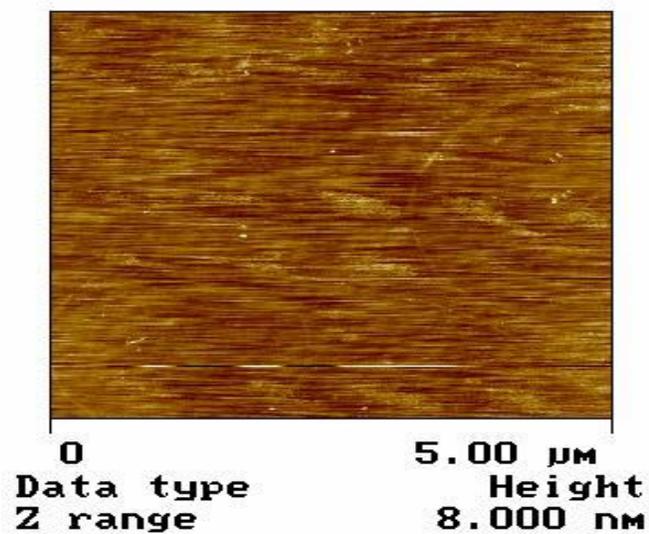

Fig2b(II): AFM image of SiC substrate

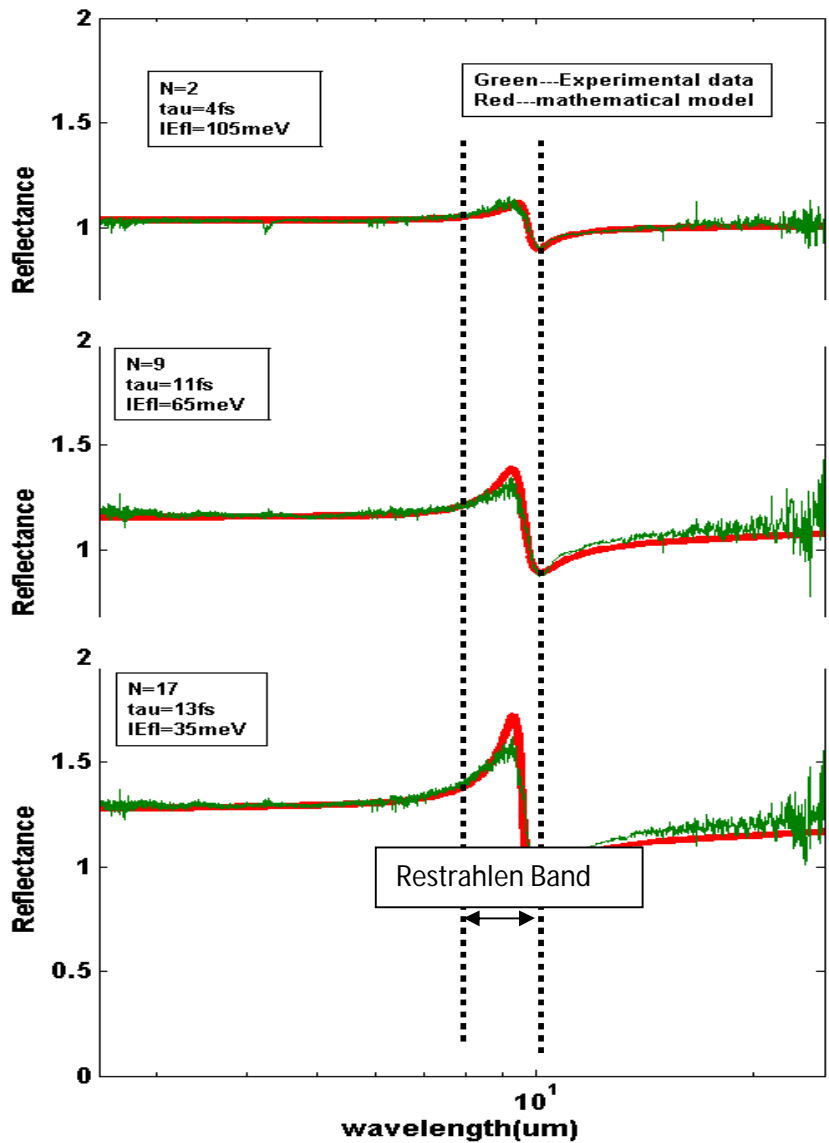

Fig3: Measured reflection spectra (green line) from 2.5μm to 25μm along with theoretical reflection spectra (red line) using eqs. (1)-(4) considering complex dielectric constant for SiC.

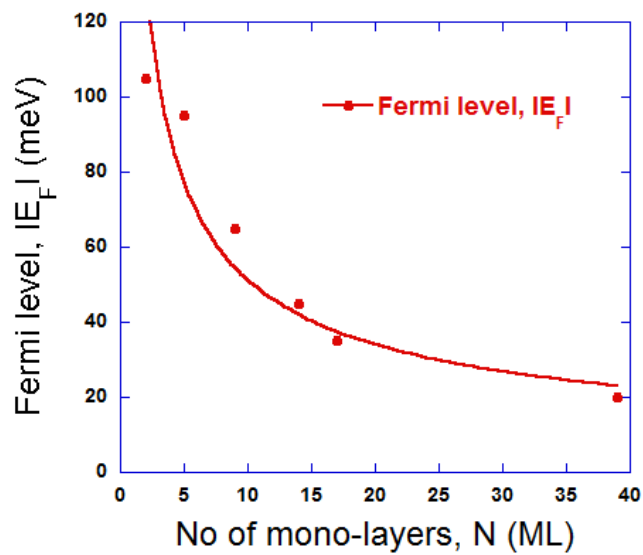

Fig4(a): Variation of |E_F| with increasing N.

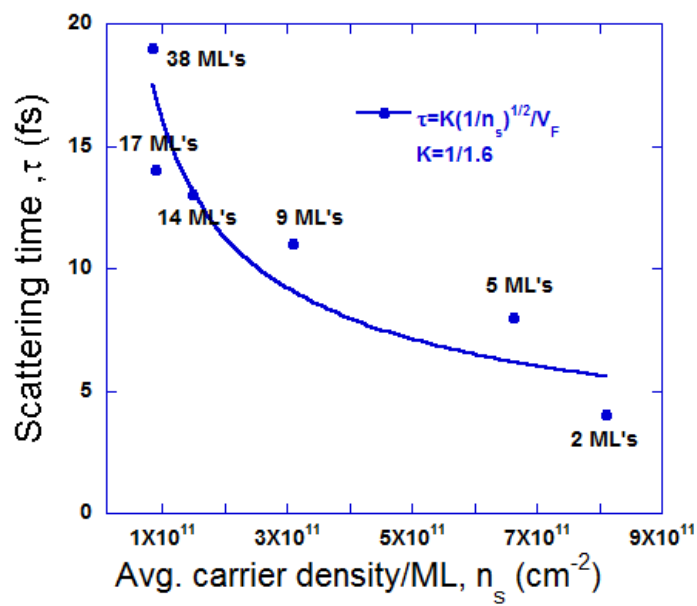

Fig4(b): Variation of τ with increasing $n_s$; increasing N has also been indicated. This τ∝1/√$n_s$ behavior indicates that short range scattering is dominant.